\documentclass[aps,prl,twocolumn,tightenlines]{revtex4}
\usepackage{graphicx}
\usepackage{dcolumn}
\usepackage{bm}
\usepackage{epsfig}
\usepackage{amsmath}
\input{epsf}
\usepackage{psfrag}
\usepackage{xcolor}
\usepackage{grffile}
\usepackage{graphicx}
\usepackage{multirow}
\usepackage{bm}
\usepackage{bbm}
\usepackage{color}
\usepackage{slashed}

\newcommand{\bea}{\begin{eqnarray}}
\newcommand{\eea}{\end{eqnarray}}

\begin{document}

\title{Polarization and quantum entanglement effects in $B^\pm_c\to J/\psi+\pi^\pm +\pi^0$ process}
\author{  Kaiwen Chen$^{1}$, Yiqi Geng$^{1,2}$\footnote{yqgeng@njnu.edu.cn}, Yichao  Jin$^{1}$\footnote{ yichao@njnu.edu.cn}, Zhicheng Yan$^{1,2}$, Ruilin Zhu$^{1,3,4}$\footnote{ rlzhu@njnu.edu.cn}}
\affiliation{
$^1$ Department of Physics and Institute of Theoretical Physics, Nanjing Normal University, Nanjing, Jiangsu 210023, China\\
$^2$ Institute of Modern Physics, Chinese Academy of Sciences, Lanzhou, Gansu 730000, China\\
$^3$ CAS Key Laboratory of Theoretical Physics, Institute of Theoretical Physics,
Chinese Academy of Sciences, Beijing 100190, China\\
$^4$ Peng Huanwu Innovation Research Center, Institute of Theoretical Physics,
Chinese Academy of Sciences, Beijing 100190, China}

\begin{abstract}
Motivated by the very recent observation of the $B^+_c\to J/\psi+\pi^+ +\pi^0$ decay using proton-proton collision data by the LHCb collaboration, we
study the four-body angular distributions and the quantum entanglement effects in the $B^+_c\to J/\psi+\pi^+ +\pi^0$ associated with $J/\psi\to \mu^++\mu^-$.
The helicity angular distributions are given in the QCD effective theory and the von Neumann entropy is obtained in $B^\pm_c\to J/\psi(\to \mu^+\mu^-)+\rho^\pm(\to \pi^\pm \pi^0)$ decay process.
\end{abstract}

\maketitle

\section{Introduction}

Beauty-charm mesons composed of two different heavy flavor quarks are yet to be fully explored compared with
charmonia  and bottomonia which are composed of two identical heavy flavors. Up to now, only three members of beauty-charm meson family, i.e., $B_c$,
 $B_c(2S)$ and $B^*_c(2S)$ are discovered in particle physics experiments~\cite{CDF:1998ihx,ATLAS:2014lga,CMS:2019uhm,LHCb:2019bem}.
 Other beauty-charm states are still missing  and waiting for discovery in experiments~\cite{Hao:2024nqb,Tao:2022qxa,Tao:2023mtw,Li:2023wgq,Li:2019tbn}.

The beauty-charm mesons below the $BD$ threshold major decay into the ground $B_c$ state via strong or electromagnetical interactions,
while the ground $B_c$ state has to weak decay. Therein the transition of the ground $B_c$ state into $J/\psi$ is very important
because the $J/\psi\to \mu^++\mu^-$ decay has clear signal at hadron collider experiments and it sheds light on the rare $B_c$ decays
and leads to the discoveries of tens of rare $B_c$ decay channels at LHCb.

Very recently, the observation of  $B_c$ decays into $J/\psi$ associated with two Pion mesons has been  reported for the first time at the LHCb experiment
using proton-proton collision data at centre-of-mass energies of 7, 8, and 13TeV corresponding to an integrated luminosity of $9fb^{-1}$. Though the
reconstruction efficiency for this channel is low, over one thousand  events are yielded using LHCb run-I and run-II data.
The measured ratio of the branching fractions of the $B^+_c\to J/\psi+\pi^+ +\pi^0$ and $B^+_c\to J/\psi+\pi^+ $ is given as~\cite{LHCb:2024nlg}
\begin{align}
{\cal R}=&\frac{\mathrm{Br}\left(B^+_c\to J/\psi+\pi^+ +\pi^0\right)}{\mathrm{Br}\left(B^+_c\to J/\psi+\pi^+ \right)}\nonumber\\
      =&2.80\pm0.15\pm0.11\pm0.16.\nonumber
\end{align}

In history, the first exclusive decay $B^+_c\to J/\psi+\pi^+ $ was discovered until 2005 by the CDF collaboration~\cite{CDF:2005yjh}, while the
multi-body exclusive decays $B^+_c\to J/\psi+\pi^++\pi^++\pi^- $ and $B^+_c\to J/\psi+3\pi^++2\pi^- $ were discovered by the CMS and LHCb collaborations in 2014~\cite{CMS:2014oqy,LHCb:2014acd}, respectively.  Then the $B_c\to J/\psi+n\pi$ decay with Pion meson numbers $n=1, 2, 3, 5$ are all observed in particle physic experiments~\cite{ParticleDataGroup:2022pth}.
The discovery of $B_c$ decays into $J/\psi$ associated with two Pion mesons with relative large ratio shall indicate the experimental feasibility of  $B_c$ decays into $J/\psi$ associated with four Pion mesons.

On the other hand, the phenomenon of quantum entanglement and the violation of Bell inequality are widely studied in quantum mechanics via electromagnetic forces. However, there are few studies of quantum entanglement and the violation of Bell inequality via strong or/and weak interactions which would tell us the essential behavior of the Quantum Chromodynamics (QCD) field theory and the standard  electroweak unified theory of particle physics.  The quantum entanglement effects in hadron decays are studied in literatures~\cite{Tornqvist:1980af,Privitera:1991nz,Abel:1992kz,Li:2006fy,Li:2009rta,Fabbrichesi:2023idl,Han:2023fci}.

In this paper, we study the polarization and quantum entanglement effects in $B^\pm_c\to J/\psi+\pi^\pm +\pi^0$ process. First, we analyze the invariant mass distribution of two Pion mesons in the $B^+_c\to J/\psi+\pi^+ +\pi^0$ decay and find that the mediate vector rho meson dominates the decay of $B^+_c\to J/\psi+\pi^+ +\pi^0$. Next we investigate the helicity angle distribution in $B^\pm_c\to J/\psi(\to \mu^+\mu^-)+\rho^\pm(\to \pi^\pm \pi^0)$ decay based on the
Jacob-Wick theory. Using the decay amplitudes in nonrelativistic Quantum Chromodynamics (NRQCD) effective theory, the helicity angle distributions are plotted. These angle distributions can be employed to test the QCD effective theory and the fundamental properties of the initial beauty-charm mesons.
Then we study the quantum spin entanglement of two vector mesons $J/\psi$ and $\rho$ in $B^\pm_c\to J/\psi+\pi^\pm +\pi^0$ process. The von Neumann entropy to describe the degree of entanglement is derived in the end.

\section{Angular distributions}

The invariant mass distribution of two Pion mesons in $B^+_c\to J/\psi+\pi^+ +\pi^0$ have been studied previously in literatures~\cite{Luchinsky:2012rk,Luchinsky:2022pxu,Geng:2023ffc} before the experimental  discovery of this channel~\cite{LHCb:2024nlg}.
Using the method in Ref.~\cite{Geng:2023ffc} but with different parametrization form, the spectral function $\rho^{2\pi}_{T}(m_{2\pi})$
can de written as
\begin{align}
            \rho_T^{2\pi}(m)=&\frac{a^3}{2m}\left(\frac{m^2 - 4m^2_{\pi}}{m^2}\right)^{2} \left(1-b\,m^2\right)
            \nonumber\\ &\times \left[\frac{1}{c^2/4+(m-m_1)^2}\right].
 \end{align}

\begin{figure}[th]
\includegraphics[width=0.45\textwidth]{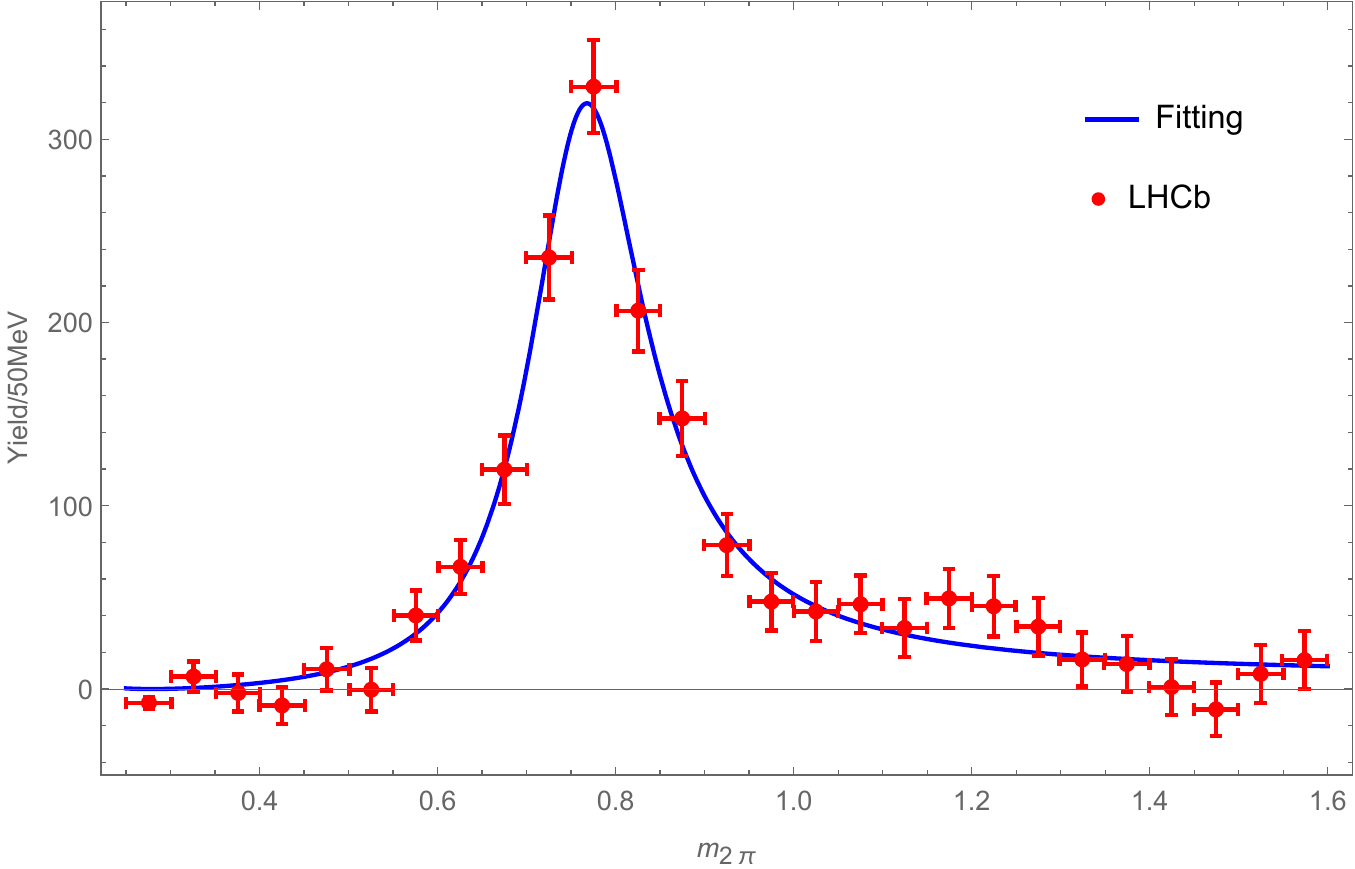}
\caption{Invariant mass distribution of $\pi^+\pi^0$ in the $B^+_c\to J/\psi+\pi^+ +\pi^0$ decay, where the LHCb data is from Ref.~\cite{LHCb:2024nlg}.}\label{fig:m2pi}
\end{figure}

Refitting the LHCb data, the parameters can be determined as $a=8.25\times10^4GeV$, $b=0.37GeV^{-2}$, $c=0.158GeV$ and $m_1=0.7518GeV$ with
$\chi^2/dof=0.96$. From the curve, the $\rho$ meson plays the dominant role in the $B^+_c\to J/\psi+\pi^+ +\pi^0$ decay. The mass and decay width of $\rho$ can be obtained through the parameter $c$ and $m_1$. The fitting curve in comparison with the LHCb data is plotted in Fig.~\ref{fig:m2pi}.

Thus we can study the dominant two-body $B^+_c\to J/\psi+\rho^+$  decay with $\rho^+\to \pi^++\pi^0$ in place of $B^+_c\to J/\psi+\pi^+ +\pi^0$ decay. The pure three-body decays without mediate resonances in $B^+_c\to J/\psi+\pi^+ +\pi^0$  are small and ignored.   The decay amplitude of $B_c\to J/\psi+\rho$ can be written as
\begin{align}
h_{\lambda}\equiv &\langle \rho(p_1, \lambda) J/\psi(p_2,\lambda)|{\cal H}_{eff}|B_c(p)\rangle\nonumber\\
=&\epsilon_{1 \mu}(\lambda)^* \epsilon_{2 \nu}(\lambda)^*\left(a g^{\mu \nu}+\frac{b p^\mu p^\nu}{m_1 m_2} +\frac{i c \epsilon^{\mu \nu \alpha \beta} p_{1 \alpha} p_\beta}{m_1 m_2} \right),
\end{align}
where the helicities satisfy $\lambda=\lambda_1=\frac{\vec{p}_{1}\cdot \vec{S}_{1}}{|\vec{p}_{1}|}=\lambda_2=\frac{\vec{p}_{2}\cdot \vec{S}_{2}}{|\vec{p}_{2}|}$ with  $\lambda\in\left(\pm1,~0\right)$ due to the conservation of momentum and angular momentum in this process.  $m_i$, $p_i$ and $\epsilon_i$ are the mass, momentum and polarization vector for  $\rho$ and $J/\psi$, respectively. $a$, $b$ and $c$ can be calculated in certain theoretical frameworks. The weak effective Hamiltonian ${\cal H}_{eff}$ governing $B_c\to J/\psi+\rho$ can be written as
\begin{eqnarray}
{\cal H}_{
eff}=\frac{G_{F}}{\sqrt{2}}V_{ud}^{*}V_{cb}\left(C_1(\mu) Q_{1}(\mu)
+C_{2}(\mu)Q_{2}(\mu)\right)\,,
\end{eqnarray}
where $V_{ud}$ and $V_{cb}$ are the
Cabibbo-Kobayashi-Maskawa (CKM) matrix elements. $Q_{1,2}(\mu)$ are
the effective four-fermion operators while $C_{1,2}(\mu)$ are the
corresponding Wilson coefficients~\cite{Qiao:2012hp}.

\begin{figure}[th]
\includegraphics[width=0.45\textwidth]{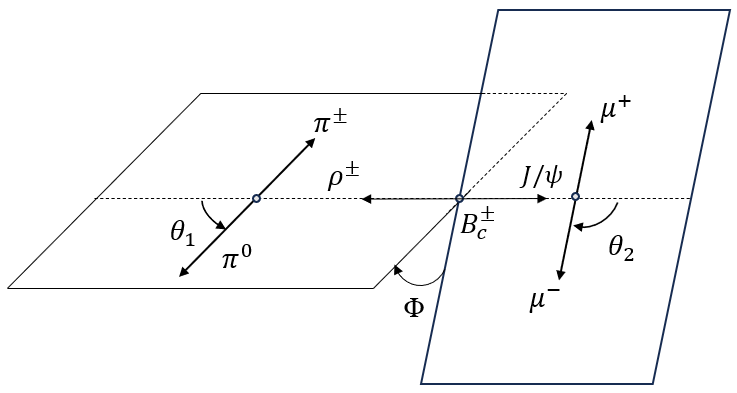}
\caption{Helicity angles in the sequential decay process $B^\pm_c\to J/\psi+\rho^\pm\to(\mu^+\mu^-)(\pi^\pm\pi^0)$,  where $\theta_1$ is the Pion meson angle in the rest of $\rho$ while $\theta_2$ is the muon angle in the rest of $J/\psi$. $\phi$ is the angle between the two subprocess decay planes. }\label{fig:angles}
\end{figure}

After the decay $B_c\to J/\psi+\rho$,  the $\rho$ further decays into two Pion mesons almost 100\% while $J/\psi$ can further decay into $\mu^+\mu^-$ with the branching ratio 5.96\%~\cite{ParticleDataGroup:2022pth}. All the decay chains are depicted in Fig.~\ref{fig:angles}, where the helicity angles $\theta_1$, $\theta_2$ and $\phi$ are defined. According to the Jacob-Wick theory~\cite{Jacob:1959at}, the angular
distributions for $B_c\to J/\psi(\to \mu^+\mu^-)+\rho (\to \pi\pi)$ can be written as

 \begin{widetext}
\begin{align}
\frac{d^3 \Gamma\left(B_c\to J/\psi(\mu^+\mu^-)+\rho ( \pi\pi)\right)}{d \cos \theta_1 d \cos \theta_2 d \phi}=\frac{9p_m}{128 \pi^2 M^2} \{ & \cos ^2 \theta_1 \sin ^2 \theta_2H_{00}+\frac{1}{4} \sin ^2 \theta_1\left(1+\cos ^2 \theta_2\right)\left(H_{11}+H_{-1-1}\right) \nonumber\\
& -\frac{1}{2} \sin ^2 \theta_1 \sin ^2 \theta_2\left[\cos 2 \phi \operatorname{Re}\left(H_{1-1}\right)-\sin 2 \phi \operatorname{Im}\left(H_{1-1} \right)\right] \nonumber\\
& \left.-\frac{1}{4} \sin 2 \theta_1 \sin 2 \theta_2\left[\cos \phi \operatorname{Re}\left(H_{10}+H_{-10} \right)-\sin \phi \operatorname{Im}\left(H_{10} -H_{-10} \right)\right]\right\},
\end{align}
 \end{widetext}
where $H_{\lambda\lambda'}=h_{\lambda}h_{\lambda'}^*$ is the density matrix elements for the $B_c\to J/\psi+\rho$ decay. $M$ is the mass of initial $B_c$ meson. The parameter $p_m$ is written as
\begin{align}
p_m^2=&\frac{m_1^2 m_2^2}{M^2}\left(\left(\frac{M^2-m_1^2-m_2^2}{2m_1 m_2}\right)^2-1\right).
\end{align}

The relationship between the parameters a,b,c and the helicity amplitude can be expressed as
\begin{align}
h_{\pm1}=a \pm \sqrt{x^2-1}c ,\nonumber \\
h_0=-ax-b(x^2-1) ,
\end{align}
where
\begin{align}
x^2=\frac{p_m^2M^2}{m_1^2m_2^2}+1.
\end{align}

\begin{figure}[th]
\includegraphics[width=0.45\textwidth]{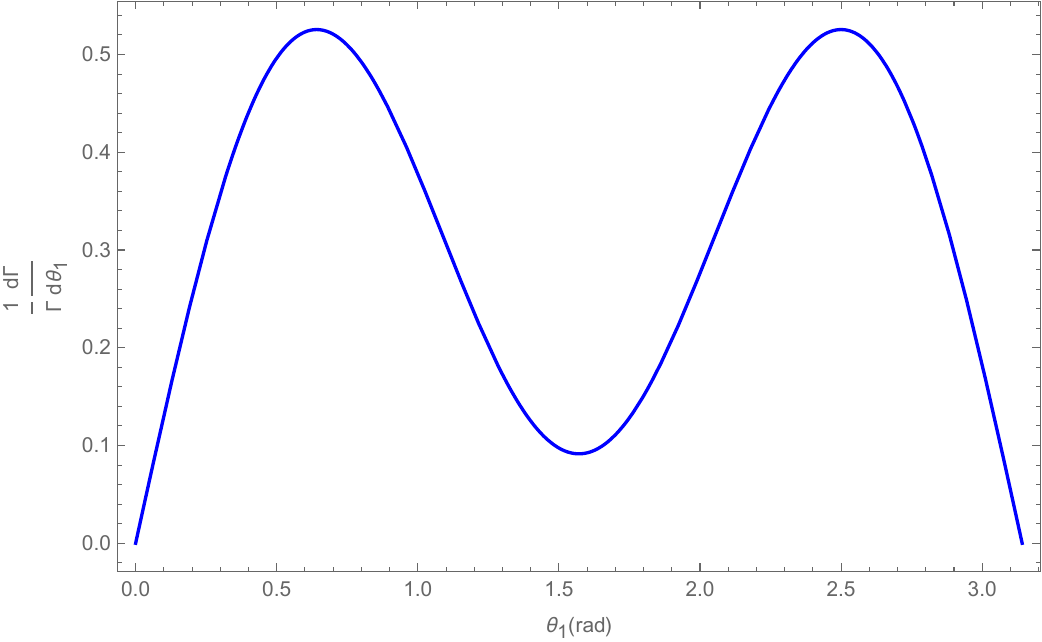}
\caption{Normalized decay width distribution with respect to helicity angle $\theta_1$.}
\label{theta1}
\end{figure}

\begin{figure}[th]
\includegraphics[width=0.45\textwidth]{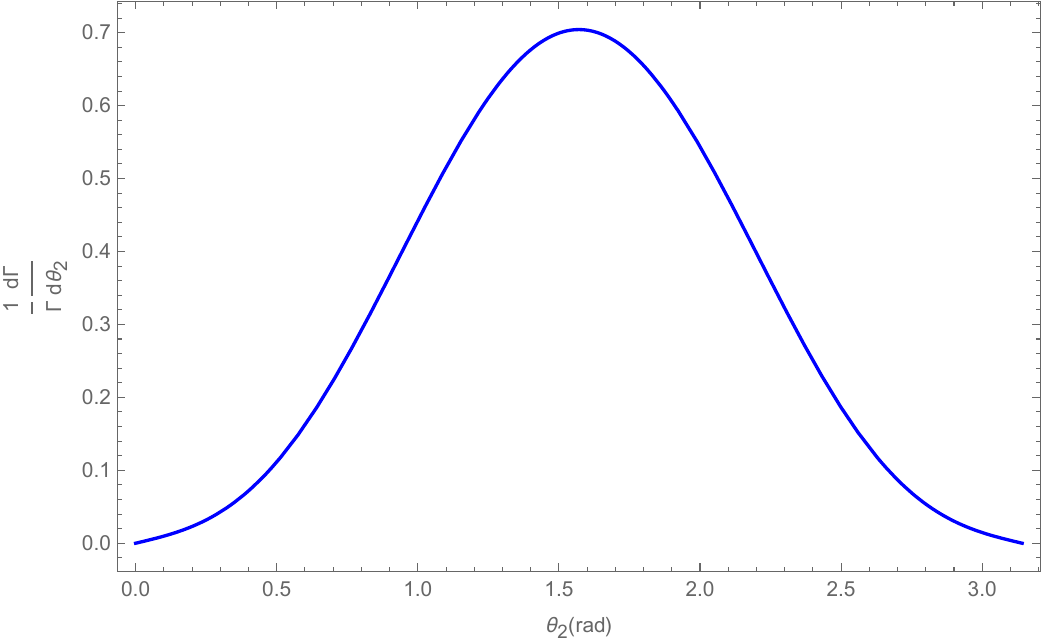}
\caption{Normalized decay width distribution with respect to helicity angle $\theta_2$.}
\label{theta2}
\end{figure}

\begin{figure}[th]
\includegraphics[width=0.45\textwidth]{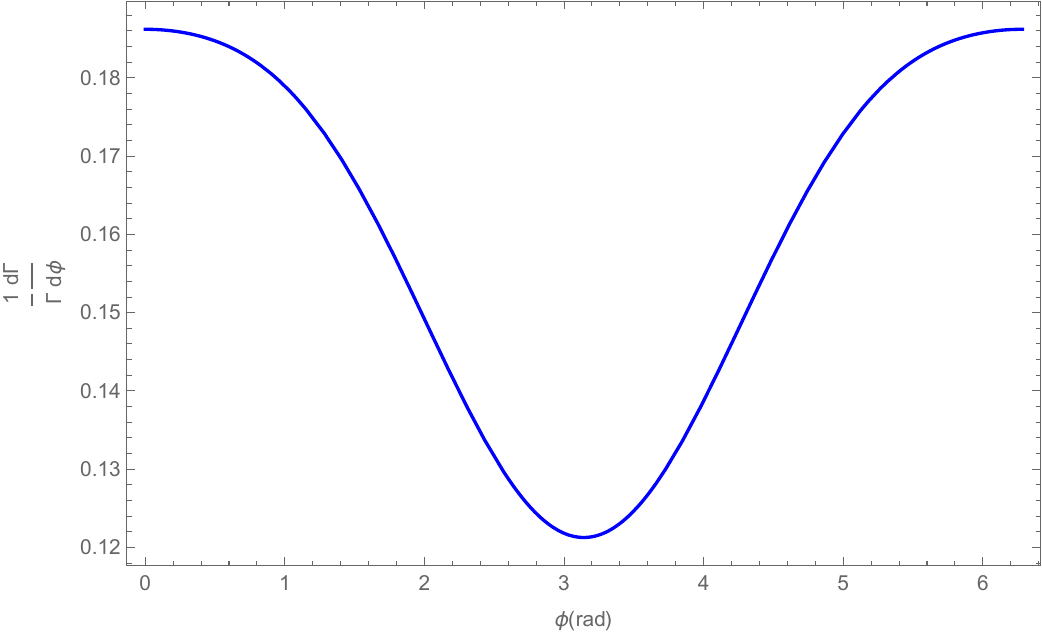}
\caption{Normalized decay width distribution with respect to helicity angle $\Phi$.}
\label{Phi}
\end{figure}

In naive factorization based on the NRQCD effective theory, the amplitude parameter $a, b, c$ can be expressed by form factors~\cite{Zhu:2017lqu,Wang:2018duy,Qiao:2012vt,Qiao:2012hp,Shen:2021dat}
\begin{align}
a=& -\frac{G_F}{\sqrt{2}}V_{ud}^{*}V_{cb}C_{1}(\mu)
f_{\rho}A_1(m_{\rho}^2)m_{\rho}(m_{B_c}+m_{J/\psi}),\nonumber\\
b=& \frac{2G_F}{\sqrt{2}}V_{ud}^{*}V_{cb}C_{1}(\mu)
f_{\rho}\frac{A_2(m_{\rho}^2)m_{J/\psi}m^2_{\rho}}{m_{B_c}+m_{J/\psi}},\nonumber\\
c=& -\frac{2G_F}{\sqrt{2}}V_{ud}^{*}V_{cb}C_{1}(\mu)
f_{\rho}\frac{V(m_{\rho}^2)m_{J/\psi} m^2_{\rho}}{m_{B_c}+m_{J/\psi}},
\end{align}
Then the values of the three helicity amplitudes can be obtained as
\begin{align}
&h_{+1}=-2.597 \times 10^{-4}MeV,  \nonumber\\
&h_{-1}=-7.794 \times 10^{-5}MeV, \nonumber\\
 &h_{0}=7.253 \times 10^{-4}MeV.
\end{align}
Using the helicity amplitudes, we plot the normalized angular distributions in Fig.~\ref{theta1}, Fig.~\ref{theta2} and Fig.~\ref{Phi}.
From these results, the distribution curves for helicity angles $\theta_1$ and $\theta_2$ are different because the two vector mesons
decay to final states with different spin structures. For the secondary $\rho^+\to \pi^++\pi^0$ decay in $B^+_c\to J/\psi+\rho^+$ process,
the maximum probability of helicity angle $\theta_1$ is around $\pi/4$, while the secondary $J/\psi\to \mu^++\mu^-$ decay in $B^+_c\to J/\psi+\rho^+$ process,
the maximum probability of helicity angle $\theta_2$ is around $\pi/2$. The secondary decays of two vector mesons refer to occur in the identical plane.

\section{Quantum entanglement and von Neumann entropy}

If the polarization state of the two vector meson system for $J/\psi$ and $\rho$ from $B_c$ is pure, the wave function 
in the rest frame of $B_c$ can be written as~\cite{Fabbrichesi:2023cev}
\begin{align}
| \Psi \rangle=&\frac{1}{\sqrt{\left\vert H \right\vert ^2}}\left[ h_{+1}|J/\psi(+1)\rho(+1)\rangle \right.\nonumber \\
 &\left.+h_0|J/\psi(0)\rho(0)\rangle+h_{-1}|J/\psi(-1)\rho(-1)\rangle\right],
\end{align}
where the $+1,~-1,~0$ in the bracket of the vector mesons denote the helicity values.
The normalized constant $H$ is defined as
\begin{align}
\left\vert H \right\vert ^2= \left\vert h_{+1} \right\vert ^2+\left\vert h_0 \right\vert ^2+\left\vert h_{-1}\right\vert ^2.
\end{align}

Then it is easy to write down the polarization density matrix $\varrho=| \Psi \rangle \langle \Psi |$ in terms of the polarization amplitude
\begin{align}
\varrho=\frac{1}{\left\vert H \right\vert ^2}
\left(
\begin{matrix}
0&0&0&0&0&0&0&0&0 \\
0&0&0&0&0&0&0&0&0 \\
0&0&h_{+1}h_{+1}^*&0&h_{+1}h_0^*&0&h_{+1}h_{-1}^*&0&0 \\
0&0&0&0&0&0&0&0&0 \\
0&0&h_0h_{+1}^*&0&h_0h_0^*&0&h_0h_{-1}^*&0&0 \\
0&0&0&0&0&0&0&0&0 \\
0&0&h_{-1}h_{+1}^*&0&h_-h_0^*&0&h_{-1}h_{-1}^*&0&0 \\
0&0&0&0&0&0&0&0&0 \\
0&0&0&0&0&0&0&0&0 \\
\end{matrix}
\right).
\end{align}
These helicity amplitudes can be calculated according to  certain theoretical frames presented in this paper.

In order to study the entanglement of two polarized vector particles with mass, the von Neumann entropy is a good choice to quantify the entanglement. The von Neumann entropy is defined as~\cite{Horodecki:2009zz}
\begin{align}
\varepsilon=-Tr[\varrho_A\ln\varrho_A]=-Tr[\varrho_B\ln\varrho_B].
\end{align}
$\varrho_A$ and $\varrho_B$ are reduced density matrices of two subsystems, which can be obtained by tracing one of the subsystems by the density operator.In this article, the two subsystems are $J/\psi$ and $\rho$. For a two-qutrit system, the von Neumann entropy satisfies $0 \leq \varepsilon \leq \sqrt{3}$.When the wave function of the two systems can be written as a direct product state, the minimum value is taken. When the entanglement of the two systems is maximum, the maximum value is taken. It is easy to figure out that the reduced density matrices of two sub systems satisfy
\begin{align}
\varrho_{J/\psi}=\varrho_{\rho}=\frac{1}{\left\vert H \right\vert ^2}
\left(
\begin{matrix}
h_{+1}h_{+1}^*&0&0 \\
0&h_0h_0^*&0 \\
0&0&h_{-1}h_{-1}^* \\
\end{matrix}
\right).
\end{align}

Combined with Eqs.~(9-11), the von Neumann entropy of subsystems $J/\psi$ and $\rho$ can be obtained
\begin{align}
\varepsilon=0.405.
\end{align}
The result of von Neumann entropy proved that there is quantum spin entanglement between two daughter vector particles in the $B_c$ decay.

\section{Conclusion}

In this paper, the angular distributions of $B^\pm_c\to J/\psi(\to \mu^+ \mu^-) + \rho^{\pm}(\to \pi^{\pm}\pi^0) $ process  are given in the helicity form of Jacob-Wick theory and the quantum spin entanglement effects among vector mesons are studied. To describe the degree of quantum spin entanglement in this process, the von Neumann entropy is calculated. Through the measurement of the invariant mass distribution of two Pion mesons, the  secondary decay angular distributions in $\rho^{\pm}\to\pi^{\pm}+\pi^0$ and  $J/\psi\to \mu^+ + \mu^-$, and the distribution with respect to the angle between the two secondary decay planes, the nature of the initial beauty-charm meson can be determined. On the other hand, these measurements of angular distributions shall provide new views to hunting for the missing members of
beauty-charm meson family. The study of the quantum spin entanglement effects in hadron cascade decays will be useful for us to understand the quantum correlations and non-locality in standard field theory of particle physics.

\section*{Acknowledgments}
  This work is supported by NSFC under grant No.~12322503, No.~12047503, and No.~12075124,
 and by Natural Science Foundation of Jiangsu under Grant No.~BK20211267.


\begin{thebibliography}{99}
\bibitem{CDF:1998ihx}
F.~Abe \textit{et al.} [CDF],
Phys. Rev. Lett. \textbf{81}, 2432-2437 (1998)
doi:10.1103/PhysRevLett.81.2432
[arXiv:hep-ex/9805034 [hep-ex]].

\bibitem{ATLAS:2014lga}
G.~Aad \textit{et al.} [ATLAS],
Phys. Rev. Lett. \textbf{113}, no.21, 212004 (2014)
doi:10.1103/PhysRevLett.113.212004
[arXiv:1407.1032 [hep-ex]].

\bibitem{CMS:2019uhm}
A.~M.~Sirunyan \textit{et al.} [CMS],
Phys. Rev. Lett. \textbf{122}, no.13, 132001 (2019)
doi:10.1103/PhysRevLett.122.132001
[arXiv:1902.00571 [hep-ex]].

\bibitem{LHCb:2019bem}
R.~Aaij \textit{et al.} [LHCb],
Phys. Rev. Lett. \textbf{122}, no.23, 232001 (2019)
doi:10.1103/PhysRevLett.122.232001
[arXiv:1904.00081 [hep-ex]].

\bibitem{Hao:2024nqb}
W.~Hao and R.~Zhu,
[arXiv:2402.18898 [hep-ph]].

\bibitem{Tao:2022qxa}
W.~Tao, R.~Zhu and Z.~J.~Xiao,
Phys. Rev. D \textbf{106}, no.11, 114037 (2022)
doi:10.1103/PhysRevD.106.114037
[arXiv:2209.15521 [hep-ph]].

\bibitem{Tao:2023mtw}
W.~Tao, Z.~J.~Xiao and R.~Zhu,
JHEP \textbf{05}, 189 (2023)
doi:10.1007/JHEP05(2023)189
[arXiv:2303.07220 [hep-ph]].

\bibitem{Li:2023wgq}
X.~J.~Li, Y.~S.~Li, F.~L.~Wang and X.~Liu,
Eur. Phys. J. C \textbf{83}, no.11, 1080 (2023)
doi:10.1140/epjc/s10052-023-12237-9
[arXiv:2308.07206 [hep-ph]].

\bibitem{Li:2019tbn}
Q.~Li, M.~S.~Liu, L.~S.~Lu, Q.~F.~L\"u, L.~C.~Gui and X.~H.~Zhong,
Phys. Rev. D \textbf{99}, no.9, 096020 (2019)
doi:10.1103/PhysRevD.99.096020
[arXiv:1903.11927 [hep-ph]].

\bibitem{LHCb:2024nlg}
R.~Aaij \textit{et al.} [LHCb],
[arXiv:2402.05523 [hep-ex]].

\bibitem{CDF:2005yjh}
A.~Abulencia \textit{et al.} [CDF],
Phys. Rev. Lett. \textbf{96}, 082002 (2006)
doi:10.1103/PhysRevLett.96.082002
[arXiv:hep-ex/0505076 [hep-ex]].

\bibitem{CMS:2014oqy}
V.~Khachatryan \textit{et al.} [CMS],
JHEP \textbf{01}, 063 (2015)
doi:10.1007/JHEP01(2015)063
[arXiv:1410.5729 [hep-ex]].

\bibitem{LHCb:2014acd}
R.~Aaij \textit{et al.} [LHCb],
JHEP \textbf{05}, 148 (2014)
doi:10.1007/JHEP05(2014)148
[arXiv:1404.0287 [hep-ex]].

\bibitem{ParticleDataGroup:2022pth}
R.~L.~Workman \textit{et al.} [Particle Data Group],
PTEP \textbf{2022}, 083C01 (2022)
doi:10.1093/ptep/ptac097

\bibitem{Tornqvist:1980af}
N.~A.~Tornqvist,
Found. Phys. \textbf{11}, 171-177 (1981)
doi:10.1007/BF00715204

\bibitem{Privitera:1991nz}
P.~Privitera,
Phys. Lett. B \textbf{275}, 172-180 (1992)
doi:10.1016/0370-2693(92)90872-2

\bibitem{Abel:1992kz}
S.~A.~Abel, M.~Dittmar and H.~K.~Dreiner,
Phys. Lett. B \textbf{280}, 304-312 (1992)
doi:10.1016/0370-2693(92)90071-B

\bibitem{Li:2006fy}
J.~Li and C.~F.~Qiao,
Phys. Rev. D \textbf{74}, 076003 (2006)
doi:10.1103/PhysRevD.74.076003
[arXiv:hep-ph/0608077 [hep-ph]].

\bibitem{Li:2009rta}
J.~Li and C.~F.~Qiao,
Sci. China Phys. Mech. Astron. \textbf{53}, 870-875 (2010)
doi:10.1007/s11433-010-0202-2
[arXiv:0903.1246 [hep-ph]].

\bibitem{Fabbrichesi:2023idl}
M.~Fabbrichesi, R.~Floreanini, E.~Gabrielli and L.~Marzola,
Phys. Rev. D \textbf{109}, no.3, L031104 (2024)
doi:10.1103/PhysRevD.109.L031104
[arXiv:2305.04982 [hep-ph]].

\bibitem{Han:2023fci}
T.~Han, M.~Low and T.~A.~Wu,
[arXiv:2310.17696 [hep-ph]].

\bibitem{Luchinsky:2012rk}
A.~V.~Luchinsky,
Phys. Rev. D \textbf{86}, 074024 (2012)
doi:10.1103/PhysRevD.86.074024
[arXiv:1208.1398 [hep-ph]].

\bibitem{Luchinsky:2022pxu}
A.~V.~Luchinsky,
Phys. Lett. B \textbf{832}, 137269 (2022)
doi:10.1016/j.physletb.2022.137269
[arXiv:2204.01136 [hep-ph]].

\bibitem{Geng:2023ffc}
Y.~Geng, M.~Cao and R.~Zhu,
[arXiv:2310.03425 [hep-ph]].

\bibitem{Qiao:2012hp}
C.~F.~Qiao, P.~Sun, D.~Yang and R.~L.~Zhu,
Phys. Rev. D \textbf{89}, no.3, 034008 (2014)
doi:10.1103/PhysRevD.89.034008
[arXiv:1209.5859 [hep-ph]].

\bibitem{Jacob:1959at}
M.~Jacob and G.~C.~Wick,
Annals Phys. \textbf{7}, 404-428 (1959)
doi:10.1006/aphy.2000.6022

\bibitem{Zhu:2017lqu}
R.~Zhu, Y.~Ma, X.~L.~Han and Z.~J.~Xiao,
Phys. Rev. D \textbf{95}, no.9, 094012 (2017)
doi:10.1103/PhysRevD.95.094012
[arXiv:1703.03875 [hep-ph]].

\bibitem{Wang:2018duy}
W.~Wang and R.~Zhu,
Int. J. Mod. Phys. A \textbf{34}, no.31, 1950195 (2019)
doi:10.1142/S0217751X19501951
[arXiv:1808.10830 [hep-ph]].

\bibitem{Qiao:2012vt}
C.~F.~Qiao and R.~L.~Zhu,
Phys. Rev. D \textbf{87}, no.1, 014009 (2013)
doi:10.1103/PhysRevD.87.014009
[arXiv:1208.5916 [hep-ph]].

\bibitem{Shen:2021dat}
D.~Shen, H.~Ren, F.~Wu and R.~Zhu,
Int. J. Mod. Phys. A \textbf{36}, no.19, 2150135 (2021)
doi:10.1142/S0217751X21501359

\bibitem{Fabbrichesi:2023cev}
M.~Fabbrichesi, R.~Floreanini, E.~Gabrielli and L.~Marzola,
Eur. Phys. J. C \textbf{83}, no.9, 823 (2023)
doi:10.1140/epjc/s10052-023-11935-8
[arXiv:2302.00683 [hep-ph]].

\bibitem{Horodecki:2009zz}
R.~Horodecki, P.~Horodecki, M.~Horodecki and K.~Horodecki,
Rev. Mod. Phys. \textbf{81}, 865-942 (2009)
doi:10.1103/RevModPhys.81.865
[arXiv:quant-ph/0702225 [quant-ph]].

\end{thebibliography}
\end{document}